\documentclass[ twocolumn,
superscriptaddress,
amsmath,amssymb,aps,showkeys,showpacs,
twoside,final,secnumarabic,
nofootinbib]{revtex4-2}

\usepackage[paperwidth=205mm,paperheight=290mm,top=17mm,bottom=25mm,
inner=17mm,outer=17mm,
twoside]{geometry}

\usepackage{cmap} 
\usepackage[T1,T2A]{fontenc}
\usepackage[utf8]{inputenc}
\usepackage[russian,english]{babel}
\usepackage{color}
\usepackage{graphicx}
\usepackage{dcolumn}
\usepackage{bm} 
\usepackage[unicode=true,colorlinks=true,linkcolor=magenta, urlcolor=blue, citecolor = blue,breaklinks]{hyperref}
\usepackage{multirow}
\usepackage{url}
\usepackage{breakurl}
\DeclareGraphicsExtensions{.eps}

\newcount\issue
\newcount\Vol
\newcount\numb
\usepackage{fancyhdr} 
\renewcommand\thesection{\arabic{section}}
\renewcommand\thetable{\arabic{table}}

\def\Vol{\textbf{80}}
\def\numb{x}
\begin{document}

\title{JOURNAL SECTION OR CONFERENCE SECTION \\[20pt]
Post-impulsive millimeter emission of the 2022-05-04 solar flare} 

\def\addressa{Central Astronomical Observatory at Pulkovo of Russian Academy of Sciences, St. Petersburg, 196140, Russia}
\def\addressb{Ioffe Institute, Polytekhnicheskaya, 26, St. Petersburg, 194021, Russia}
\def\addressc{Crimen Astrophysical Observatory, Nauchny, 298409}
\def\addressd{ITMO University, St. Petersburg, 197101, Russia}

\author{\firstname{G.G.}~\surname{Motorina}}
\email[E-mail: ]{g.motorina@yandex.ru }
\affiliation{\addressa} 
\affiliation{\addressb}
\author{\firstname{Yu.T.}~\surname{Tsap}}
\affiliation{\addressc}
 \author{\firstname{V.V.}~\surname{Smirnova}}
\affiliation{\addressc}
\author{\firstname{A.S.}~\surname{Morgachev}}
\affiliation{\addressa}
\author{\firstname{A.S.}~\surname{Motorin}}
\affiliation{\addressd}

\received{01.08.2025}
\revised{xx.xx.2025}
\accepted{xx.xx.2025}

\begin{abstract}
The present work aims at analyzing the nature of millimeter (mm) emission observed during the post-impulsive phase of the solar flare SOL2022-05-04T08:45 (M5.7), detected by the RT-7.5 radio telescope of the Bauman Moscow State Technical University at 93 GHz. 
We investigate the relationship of mm and extreme ultraviolet (EUV) emission with variations in the temperature and coronal plasma emission measure obtained from SDO/AIA and GOES data. The results show that the enhanced mm emission at the post-impulsive phase of the flare coincides with the increase of EUV emission, indicating a connection with moderately hot ($\sim$1 MK) plasma. Based on the calculation of the differential emission measure, we determine the parameters of the post-impulsive flare plasma and conclude that the optically thin coronal plasma may contribute of about 20\% to the mm emission. 
\end{abstract}

\par  
\keywords{Sun: Flares - Sun: X-rays, EUV, Radio emission   \\[5pt]}

\maketitle
\thispagestyle{fancy}


\section{Introduction}\label{intro}

Recent ALMA (Atacama Large Millimeter/submillimeter Array) observations have opened a new window for solar studies, offering high-resolution imaging of the Sun at millimeter (mm) and submillimeter wavelengths.
Solar ALMA observations \cite{1, 2} have allowed to obtain the structure of the upper solar atmosphere in the previously inaccessible mm band.  It has been shown (see e.g., \cite{2} ) that images of the transition region and the upper chromosphere observed in the mm and ultraviolet (1600 \AA) bands exhibit similar magnetic structures, which confirms that significant fraction of mm emission comes from the chromosphere \cite{3}.

The mm emission of the quiet chromosphere is determined by the thermal bremstrahlung emission mechanism. The mm observations of active regions have shown the positive spectral slope (spectral flux increases with frequency)
 between 212 and 405 GHz \cite{4}, which has been attributed to thermal bremsstrahlung from optically thick sources with 0.6–2.9$\times10^4$ K effective temperatures.
However, the origin of mm emission with a positive spectral slope from solar flares remains unsolved \cite{5, 6, 7, 8}. 
Multiple competing mechanisms to explain the observed mm frequency-rising spectral component from solar flares have been proposed \cite{9, 10, 11, 6, 12, 13, 14, 7}. Yet, no single mechanism explains all mm flare observations. The answer likely involves multiple processes (e.g., thermal emission from heated plasma plus non-thermal electrons) with dominance shifting during flare phases. 

As for the sources of mm radiation, previously \cite{15} with FLARIX code modeled the response of the chromosphere to an electron beam bombardment during solar flares and showed that main contribution to the mm emission comes from the chromospheric and transition region plasma with temperatures of $10^4-10^5$ K.
At the same time, \cite{16} based on ALMA observations of  solar microflares at 1.2 and 3 mm with spatial resolutions of about 28" and 60" respectively, identified flare sources of mm radiation associated with hot coronal loops observed by SDO/AIA. Still it is unclear 
whether cold or hot plasma makes the dominant contribution to mm emission.
Mm sources can correspond to the footpoints or looptops of hot coronal post-flare loops. This means that both hot and cold plasma might be a source of mm flare emission at least for small-scale flare events and especially for a post-impulsive phase, when hot and cold flare plasma can make a significant contribution to the mm radio emission.

The present work aims at the study of a standart M5.7-class solar flare SOL2022-05-04T08:45, recorded by the RT-7.5 radio telescope of the Bauman Moscow State Technical University at 93 GHz based on the multiwavelength observations and investigation of the origin of mm emission observed during its post-impulsive phase.

\section{OBSERVATIONAL DATA AND PROCESSING}\label{Section2}

\begin{figure*}
\includegraphics[width=1.08\textwidth]{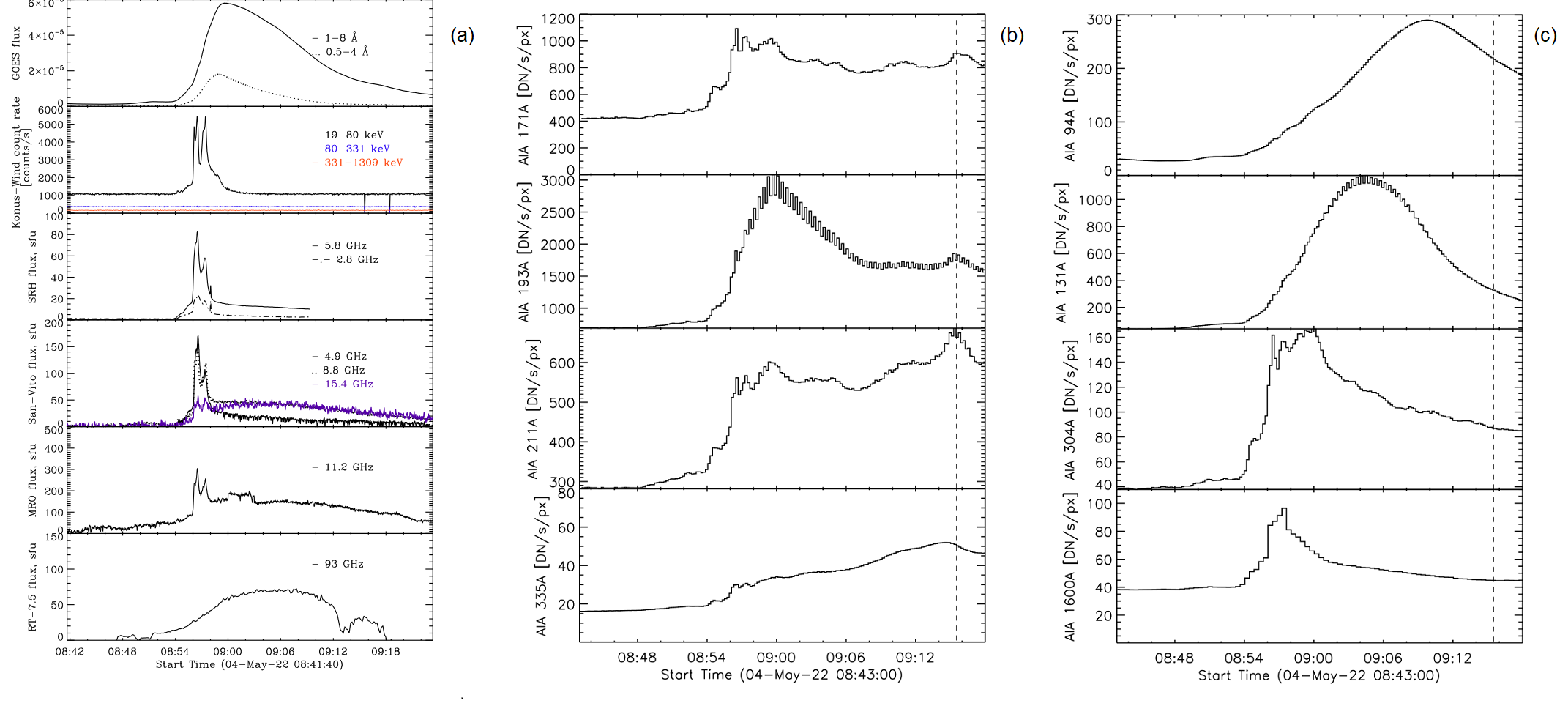}
\caption{\label{fig1} 
Light curves of SOL2022-05-04T08:45. Panel (a), from top to bottom: 1-8 and 0.5-4 \AA\; (GOES); 20–80 keV (Konus-
Wind); 2.8 and 5.8 GHz (SRH); 4.9, 8.8, 15.4 GHz (RSTN, San-Vito), 11.2 GHz (Metsähovi); 93 GHz (3.2 mm) (RT-7.5); 
panel (b): 171, 193, 211, 335 \AA\; from the ROI, shown in Fig.~\ref{fig3} (right panel); panel (c): 94, 131, 304, 1600 \AA\; from the ROI.
}
\end{figure*}

Active region NOAA 13004 (S15, W16) produced a M5.7-class confined SOL2022-05-04T08:45 solar flare, which began at 08:45:00 UT and reached its maximum intensity at 08:59:00 UT (Fig.~\ref{fig1}) according to GOES \cite{17}. The flare was well observed using multiple space- and groundbased instruments and was studied in detail in \cite{18, 19} . Fig.~\ref{fig1} shows the time profiles of the 
flare in X-rays (GOES), hard X-rays (Konus-Wind, \cite{20, 21}), microwave (Siberian Radioheliograph, SRH, \cite{22}); RSTN, San-Vito; Metsähovi, \cite{23}), mm (RT-7.5 radio telescope), and EUV (SDO/AIA, \cite{24}) emission. 

\begin{figure*}
\includegraphics[width=0.4\textwidth]{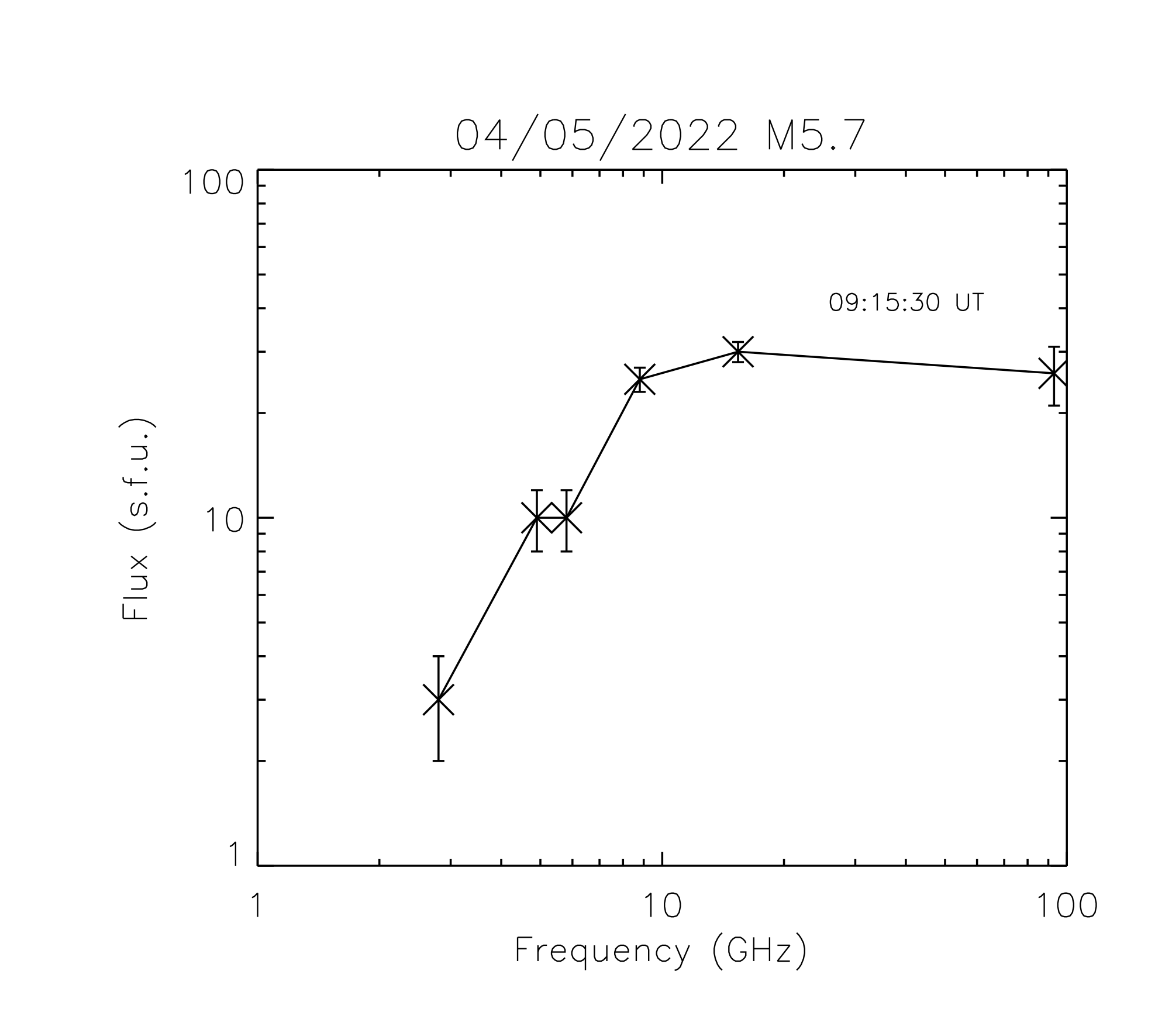}
\caption{\label{fig2} 
The radio flux density spectrum of the flare at the peak time (09:15:30 UT).}
\end{figure*}

Here we focus at its post-impulsive phase after 09:13~UT, when a noticeable increase of the mm emission at 93 GHz is observed with a peak at 09:15:30 UT (Fig.~\ref{fig1}, panel (a)). The radio flux density spectrum obtained at the peak time (09:15:30 UT) is shown in Fig.~\ref{fig2}. It can be seen that the spectrum is almost flat (spectral index between 93 and 15.4 GHz is $-0.06$) at mm frequencies, which argues in favor of the thermal free-free mechanism from the optically thin source. However, a definitive conclusion is difficult because of insufficient observations in the mm wavelength range.

\begin{figure*}
\includegraphics[width=0.78\textwidth]{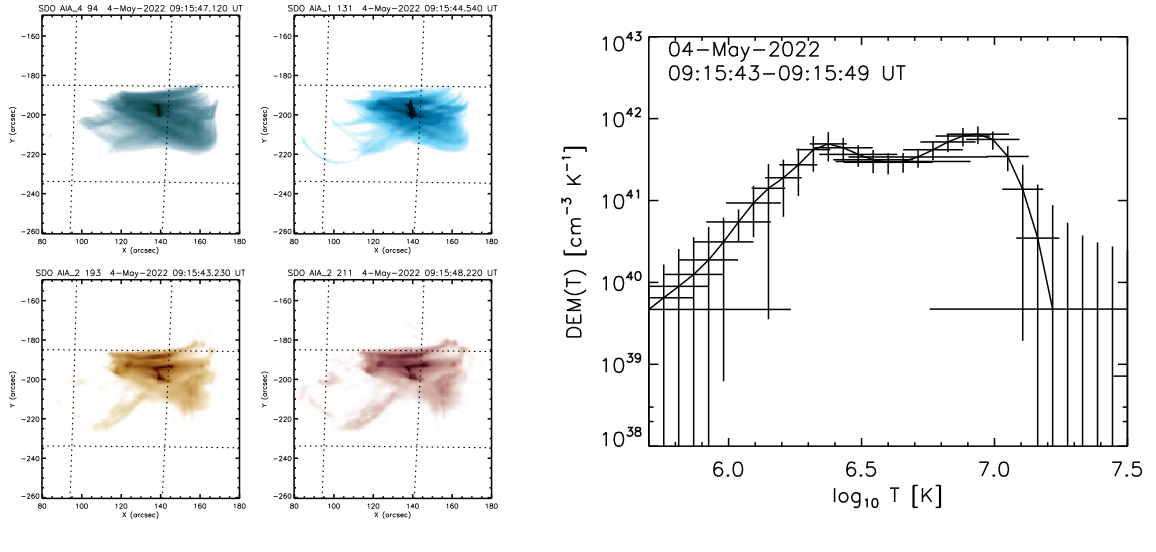}
\caption{\label{fig3} Left panel: SDO/AIA maps indicating the region of interest (ROI) taken during the post-impulsive phase of the 2022 May 4 flare. Right panel: DEM from the ROI close to the peak time (09:15:30 UT).}
\end{figure*}

A similar peak at 09:15:30 UT is observed in several EUV passbands. Fig.~\ref{fig1}, panel (b) shows the increase in EUV emission in 171, 193, 211, 335 \AA\;channels, inferred from the region of interest (ROI), shown in Fig.~\ref{fig3}. In contrast, such behavior is not observed in 94, 131, 304, 1600 \AA\; bands. According to SDO/AIA response functions \cite{24} the passbands are most sensitive to plasma temperatures as follows: 0.09 MK (304 \AA), 0.8 MK (171 \AA), 1.4 MK (193 \AA), 1.8 MK (211 \AA), 2.5 MK (335 \AA), and 7.1 MK (94 \AA). 
The increase in EUV intensity, similar to rise seen at 93 GHz, is observed only in the 171, 193, 211, and 335 \AA\;bands, which correspond to moderately warm plasma with T$\sim1$ MK. This means that EUV plasma with T$\sim1$ MK can contribute to the generation of mm emission.  

\begin{figure*}
\includegraphics[width=0.78\textwidth]{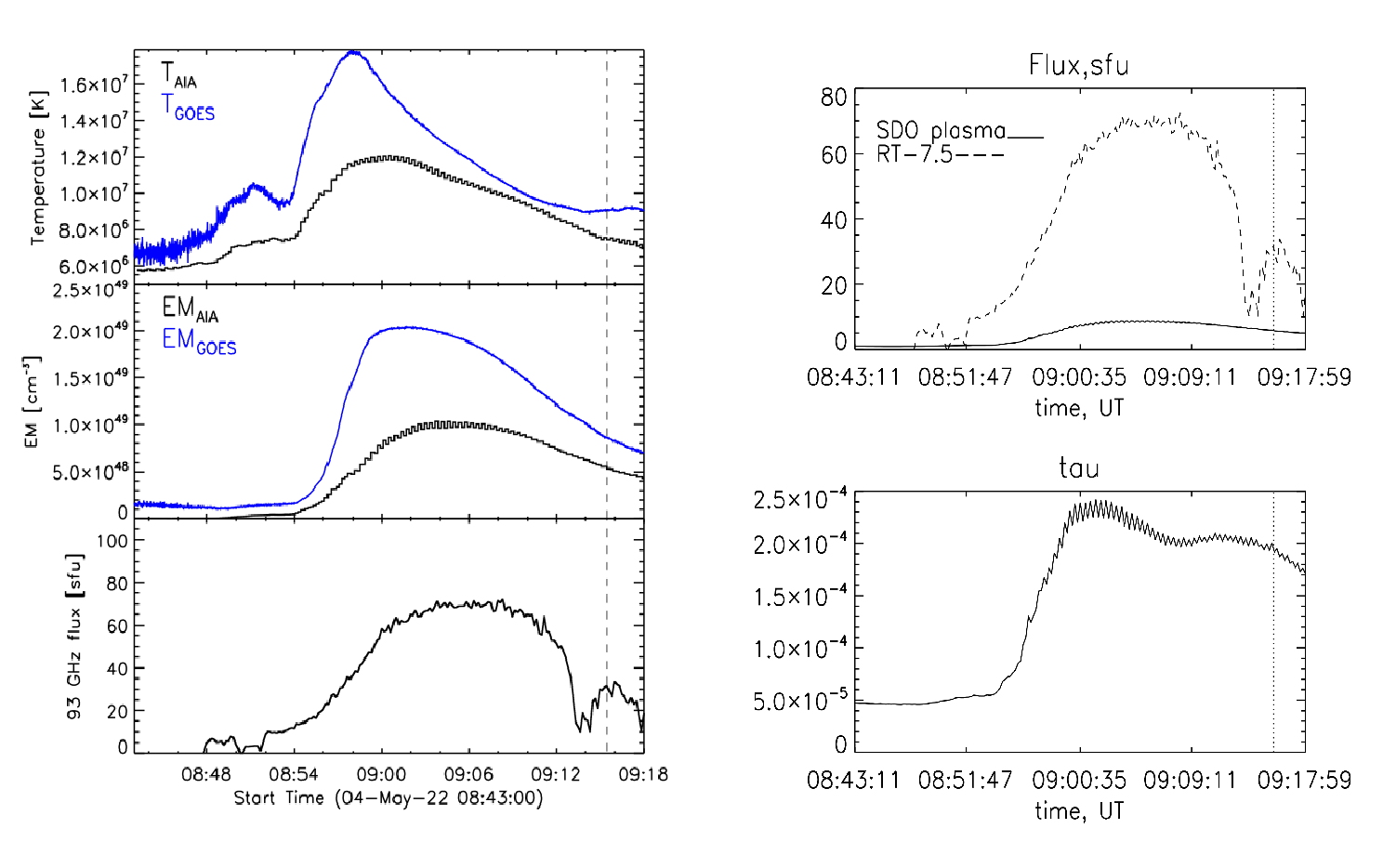}

\caption{\label{fig4} 
Comparison of the temporal profiles of SOL2022-05-04T08:45: EM, T, and mm emission (left panel); 
calculated (SDO plasma) and observed (RT-7.5) mm emission at 93 GHz and optical thicknes (right panel).}
\end{figure*}

The differential emission measure (DEM, Fig.~\ref{fig3}, right panel) from the ROI (see Fig.~\ref{fig3}) of the flare source with the area of $5.84\times10^{19}$ [cm$^2$] was recovered for 0.5–25 MK temperature range employing the methodology and numerical routines developed by \cite{25}. 
We then calculated the ROI-averaged emission measure (EM) and temperature (T) from the ROI, shown in Fig.~\ref{fig4} (left panel) and compared with EM and T obtained from the GOES data as well as with mm radio flux at 93 GHz. The results demonstrate that neither EM nor T show any increase during the post-impulsive phase.

The previously obtained DEM distributions were then used to compute the thermal bremsstrahlung emission from plasma within the specified temperature interval at 93 GHz (e.g., \cite{13}). The comparison of the observed and calculated radio flux at 93 GHz and also the optical thickness are shown in Fig.~\ref{fig4}, right panel. 
The results show that the input of the coronal plasma to mm emission is not decisive during the flare evolution. At the same time, the calculated flux at 93 GHz equals 6 [sfu], while the observed flux was $26\pm3.9$ [sfu] during the post-impulsive phase at the peak time (09:15:30 UT). This means that flare plasma can contribute significantly (up to $\sim 20 \%$) to the observed 93 GHz mm emission if the source area is large enough.

\section{DISCUSSION OF RESULTS AND CONCLUSIONS}

We have investigated how coronal flare plasma contributed to generation of the mm emission during the post-impulsive phase of the SOL2022-05-04T08:45 solar flare. 
During its post-impulsive phase (peak at 09:15:30 UT) an increase in mm and EUV emission was observed in four channels (171, 193, 211, 335 \AA), which corresponds to a moderately warm plasma with T$\sim10^6$ K.
The contribution of coronal plasma (with areas of $\sim10^{19}$ cm$^2$) to mm radiation in the post-impulsive phase of the flare is significant ($\sim$20$\%$), but not decisive.
It seems that only in the case of microflares mm radiation may originate from coronal sources. 

A comparison of the time profiles of temperature, emission measure, extreme ultraviolet, and mm emission, as well as the corresponding SDO/AIA images indicates a possible significant contribution of coronal rain to the mm component. This suggests that mm observations can be used to diagnose the process of coronal condensation during the post-mpulsive phase of the flare energy release.

In summary, we conclude that the contribution of the SDO/AIA plasma to generation of the mm emission is more significant during the post-impulsive phase of the flare, as the relative difference between the calculated flux and the observed flux of mm emission during its maximum is smaller. Consequently, the contribution of the coronal plasma increases toward the end of the flare, which agrees well with the temporal profiles of millimeter emission.


\begin{acknowledgments}
This work was supported by the  State Assignment No.FFUG-2024-0002 (GM), RSF grant No.22-12-00308-P (YT), and the State Assignment No.122022400224-7 (VS).

\end{acknowledgments}


\section*{CONFLICT OF INTEREST}
The authors of this work declare that they have no conflicts of interest.


\end{document}